\documentclass[journal]{IEEEtran}

\usepackage{cite}
\usepackage{comment}
\usepackage{bm}
\usepackage{graphicx}
\graphicspath{{./Figures/}}
\DeclareGraphicsExtensions{.pdf,.jpeg,.png}
\usepackage{amsmath}
\usepackage{amssymb}
\usepackage{algorithm}
\usepackage{algorithmic}
\usepackage{xcolor}
\usepackage{url}
\usepackage{array}
\usepackage{fancyhdr}

\usepackage{caption}  
\usepackage{subcaption}

\newcolumntype{M}[1]{>{\centering\arraybackslash}m{#1}}

\begin{document}

\title{
The most frequent N-k line outages occur in  motifs that can improve
contingency selection 
}
\author{Kai~Zhou,~\IEEEmembership{Member,~IEEE}, Ian~Dobson,~\IEEEmembership{Fellow,~IEEE,} and~Zhaoyu~Wang,~\IEEEmembership{Senior Member,~IEEE}
%\author{Kai~Zhou,~Electric Power Group \qquad Ian~Dobson and~Zhaoyu~Wang,~Iowa State University
\thanks{\looseness=-1 K. Zhou is with the School of Mechanical and Electrical Engineering, Soochow University, Suzhou, China (email: kzhou@suda.edu.cn).

I.~Dobson and Z. Wang are with the Department of Electrical and Computer Engineering, Iowa State University, Ames IA USA (email: dobson@iastate.edu; wzy@iastate.edu). 

I.~Dobson gratefully acknowledges support from NSF grant 2153163.}
}

\maketitle

\begin{abstract}
Multiple line outages that occur together show a variety of spatial patterns in the power transmission network. Some of these spatial patterns form network contingency motifs, which we define as the patterns of multiple outages that occur much more frequently than multiple outages chosen randomly from the network. 
We show that choosing N-k contingencies from these commonly occurring contingency motifs accounts for most of the probability of multiple initiating line outages.
This result is demonstrated using historical outage data for two  transmission systems.
It enables N-k contingency lists that are much more efficient in accounting for the likely multiple initiating outages than exhaustive listing or random selection.
The \mbox{N-k} contingency lists constructed from motifs can improve risk estimation in cascading outage simulations and help to confirm  utility contingency selection.
\end{abstract}
\IEEEpeerreviewmaketitle

\begin{IEEEkeywords}
Cascading risk, N-k, contingency selection, network motif,
\end{IEEEkeywords}

\section{Introduction}
\looseness=-1
It is routine to choose initial line outage contingencies to assess power transmission system security with simulation. Single line contingencies, known as N-1, are tractable and their impact is tested simply by applying each outage in turn. 
This paper analyzes the probabilities of the more challenging 
N-k initial line contingencies with k$>$1 lines outaged at once.
These multiple line contingencies, generally of higher impact and lower frequency, do occur in practice, and are simulated when assessing the risk of more extreme events such as cascading, or ensuring robustness to a list of contingencies that goes beyond N-1. We now explain how these applications motivate our analysis of the probability of N-k initial outages based on outage data routinely collected by utilities.

\looseness=-1
When assessing the risk of cascading outages with a simulation, it is usual to sample multiple initial line outages with some sort of equal probability assumption \cite{camsPS12,carrerasHICSS13}, such as independent, equal probabilities for the individual line outages that make up each multiple outage, or equal probabilities for all \hbox{N-2} outages. 
These equal probability assumptions are pragmatic
%(accurately estimating individual line outage probabilities even in unstressed conditions is quite hard \cite{kaiPS21}), 
but unrealistic, and this systematic skewing of  the sampling towards contingencies that are unlikely in practice makes the resulting risk estimates less credible.
We use contingency motifs to sample the multiple contingencies that have significantly higher probability. When this improved sampling of initial contingencies is combined with the simulated cascading impact, better estimates of cascading risk can be obtained.

The power industry routinely tests and maintains power grid reliability by simulating the impacts of  a list of credible initial outage contingencies. 
The credible contingencies include all single line contingencies and a judicious selection of the huge number of multiple contingencies that are theoretically possible.
As summarized in the literature review, the credible multiple contingencies are largely chosen by their impact or by engineering judgment. 
The contingency motifs give sets of multiple outages that have a much higher probability based on the historic outage data.
Quantifying the extent to which these sets of multiple outages have occurred much more frequently in the past can help confirm and augment the contingency list. 

\looseness=-1
This paper originates from the observation from real outage data that multiple contingencies occur much more frequently in spatial patterns that we call contingency motifs. We illustrate this idea for the simplest case of N-2 outages. For N-2 outages, the contingency motif is all the outages of two lines that share a common bus; that is, they have the spatial pattern \includegraphics[height=9pt]{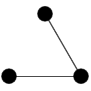}. In our first power grid example of N\,=\,528 lines, there are N(N-1)/2\,=\,139\,128 possible double contingencies, but only 2116 of these double contingencies are the contingency motif. If we assume each double contingency occurs with equal probability, then the probability of a contingency motif occurring would be 2116/139\,128\,=\,1.5\%.
However, observing the system for 14 years, we find that 81\% of the double contingencies that actually occurred are the contingency motif \includegraphics[height=9pt]{s21.pdf}. 
Since the contingency motif is much more likely to occur, we can efficiently capture much more of the probability of the realistically occurring double initial outages with  the contingency  motif than by random selection from all double outages. 

On reflection, it is not surprising that the contingency motif of two lines with a common bus occurs much more often, since the status of the two lines can be linked by both proximity and various details of the protection system and substation layout. But we can generalize this insight to N-3 and N-4 and quantify it statistically in order to capture most of the probability of N-k outages. 
The reason is that N-3 and \hbox{N-4} do occur in practice and their outage data is enough for making statistically meaningful inference; however, N-k for k$>$4 is rare and they only account for a small portion of multiple line outages, and thus there is not enough historical data for statistical analysis. 
%In particular, we analyze historical outage data from two large North American utilities.
%Bonneville Power Administration (BPA) in detail throughout the paper and apply the same analysis method to  the New York Independent System Operator (NYISO) system at the end. 

This paper 
\begin{itemize}
\item defines contingency motifs of the power network and finds that multiple initial line outages occur much more frequently in contingency motifs. 
\item develops a probabilistic model and sampling schemes for multiple contingencies.
%, showing the distribution of multiple line outages does not change significantly over the years for specific power systems.
\item  shows that the new sampling schemes and contingency lists account for most of the probability of multiple initial line outages.
\item applies to standard utility data, and analyzes historical contingencies from two large North American utilities.
\end{itemize}

\section{Literature review}
Researchers have proposed various model-based methods for contingency selection. 
The Performance Index (PI) method approximates an index for a contingency that reflects the impact on the violation of line flows or voltages \cite{wood13}. 
Then contingencies are ranked based on PIs, and those having large PIs are put on the contingency list. 
The PI method has a trade-off between approximation accuracy and computation speed. For improving approximation accuracy, \cite{polymeneasPS16} forms a PI based on three margin indices for unbalanced systems in terms of currents, voltages, and reactive power. The margin indices use the deviation beyond limits instead of the absolute values. \cite{liPS16} proposes two PI-based methods considering distribution networks. For speeding up computation, \cite{kaplunovichPS16} presents an algorithm of fast N-2 contingency selection. Based on the linear power system model, it derives a set of constraints that describe the line flow overload; whenever some contingencies are considered to be credible, the algorithm constructs new constraints to identify more credible contingencies.

Besides the PI method, some works \cite{ciapessoniSG16,  huangISGT21, weckesserVanISGT17}  first select individual critical components and then extend them to multiple contingencies. The final list is generated by screening for risky contingencies based on topological metrics or time-domain simulation. \cite{ciapessoniSG16} describes a security assessment tool that incorporates a probabilistic contingency selection module. 
This method identifies critical individual components according to the conditional probabilities of individual component outages given different threats (such as bad weather, environment, aging, sabotage). 
It then enumerates single and multiple contingencies based on critical individual components and computes topological metrics to screen for contingencies. 
Instead of conditional probabilities, \cite{huangISGT21} uses a metric based on Line Outage Distribution Factors to select critical individual components. Then, line candidates for multiple contingencies are those within a specified distance from the selected critical individual components, and multiple contingencies are ranked according to the betweenness centrality. 
Compared to the above two methods, \cite{weckesserVanISGT17} extends pre-selected contingencies only when the post-contingency state is stable, and candidates are the components that are impacted most by the pre-selected contingencies. This method uses time-domain simulation to evaluate the system voltage stability. 
\cite{McCalleyPS05} proposes a method of forming an N-k contingency list based on substation configurations. The idea is that the protection system forms functional groups, in which components outage together because of bus configurations and protection schemes. 

Moreover, statistical sampling and optimization are proposed to select multiple contingencies. \cite{eppsteinHinesPS12} proposes a Random Chemistry sampling to identify large collections of multiple contingencies that initiate cascading outages.
Specifically, Random Chemistry starts with a relatively large random subset of components that cause cascading outages, and then reduces the size of that subset recursively until a minimal subset is found. This minimal subset is a multiple contingency that leads to cascading, and it is minimal because any subset of it does not cause cascading. In a different approach, \cite{kaplunovichTuritsynHICSS14} studies the statistical properties of critical N-2 contingencies in terms of their locations in the power network, which can be used to identify critical lines.   
Mixed-integer linear programming is also used to identify credible contingencies. The objective is to maximize either the risk \cite{dingPS16} or the incremental risk \cite{WangIJEPES20} of contingencies, and a recursive algorithm is used to select a list of credible contingencies. However, the optimization method is inadequate to generate a large collection of contingencies in a limited time. \cite{sarkar22IET} formulates a mixed-integer non-linear programming problem to identify multiple contingencies that cause a large load shed. Two algorithms using power flow sensitivity and a topological metric reduce the search space to speed up computation. 

\looseness=-1
Instead of assessing cascading risk, one can pose a different question 
that starts from a large blackout caused by a very large contingency 
and then asks: what is the minimal multiple initial contingency that causes the large blackout?
This question is addressed by a Random Chemistry algorithm in\cite{eppsteinHinesPS12}.

Industry practice for contingency selection requires all single contingencies and some of multiple contingencies.
The North American Electric Reliability Corporation (NERC) established a standard TPL-001-4
\cite{nercstandard} about categories of contingencies to be adopted in transmission system planning. \cite{jiangNAPS21} discusses in detail and models the seven categories of contingencies in TPL-001-4. A conventional continuous Markov Chain is used. Its parameters, such as failure and repair rates, are estimated from outage data collection systems, such as the Transmission Availability Data System (TADS). The model evaluates the probabilities of different categories of contingencies; it does not consider specific contingencies in each category.     
\cite{robakEPE17} also discusses the standard as well as the practice of contingency analysis. It presents the experience of selecting multiple contingencies for analysis in power system planning. 
ISO New England \cite{hongMaslennikovPESGM21} is developing a tool that  calculates probabilities of multiple contingencies given weather conditions. Multiple contingencies are constructed from independent single contingencies. These single contingencies either have high probabilities or are selected by operators because they are in major power grid interfaces. 

Commercial software, such as TRELSS \cite{trelss} and PSSE \cite{psse}, have user-specified contingencies and automatic multiple contingency selection modules. User-specified contingencies could have common-mode contingencies, protection control group contingencies, and other specified contingencies. 
The automatic multiple contingency selection assumes independent individual outages in a multiple contingency, and the PI method is used for selection. 
For example, in the case of an N-2 contingency, the first outage is enumerated and ranked according to PI, and the second outage is enumerated and ranked according to PI in a subnetwork without the first outaged component; then, a combination of the first outage and the second outage is formed as an N-2, and the rank of N-k contingencies is determined by the PI of the first outage and then the PI of the second outage.      

The definition of multiple contingencies varies in different contexts: the meaning of k in N-k is different. NERC considers both primary and secondary devices, and k is the number of outaged devices. 
On the other hand, in the definition of PSSE and \cite{hongMaslennikovPESGM21}, N-1 could also be a multiple contingency under its protection scheme, which is a protection control group in \cite{McCalleyPS05}. However, \cite{McCalleyPS05} considers this contingency as an N-k, where k is the number of outaged circuits. In this paper, N-k represents a contingency involving k transmission lines.

As utilities are routinely recording outage data, data-driven and probabilistic methods for contingency selection are possible and promising but are not studied as much. One data-driven approach \cite{kaiPS21} proposes a Bayesian hierarchical model to estimate outage rates of individual transmission lines considering line dependencies. Expert knowledge is of course distilled from the experience of real outages, but there is an opportunity for statistical analysis 
to not only confirm and quantify the expert knowledge but also reveal more hidden findings that are not easily learned from experience. Motivated by this opportunity, in this paper we analyze real outage data to find historical contingency patterns and propose systematic sampling schemes for multiple contingency selection. 

It is proven in cascading simulation that initial outage spatial correlation has a substantial impact on assessing cascading risk \cite{clarfieldEppsteinHinesPSCC18,clarfeldHinesPS19}. In general, the increased correlation of close initial outages increases the cascading risk. This finding also motivates us to examine initial outage spatial patterns in real outage data.

%A straightforward probabilistic contingency selection is to estimate probabilities of contingencies and rank them accordingly. However, the problem is that there is always no sufficient outage data. Even individual contingencies are rare, not to mention multiple contingencies.  %The operation and weather conditions are changing, and states of power system components vary because of aging, maintenance, and upgrading. Therefore, 
%Besides, the probability changes over time, and we only observe a contingency under some conditions for a limited time.   

%An effective method of mitigating this problem is to group contingencies according to some common features, such as voltage ratings, lengths, component types, and temporal and spatial characteristics. 
By analyzing outage data recorded over ten years in two large power transmission systems, we find that multiple line outages occur more frequently in some spatial patterns. 
%making it an effective grouping criterion. 
This idea is inspired by the network motif concept. 
Network motifs are recurrent and statistically significant subgraphs of a network that are first introduced by complex network and biology researchers to analyze gene regulation networks \cite{alonNG02,MiloScience02}. 
Network motifs are widely used in gene regulation networks in systems biology and successfully applied in ecological, sociological, and epidemiological networks \cite{stoneCB19}. 

There are studies applying network motifs to power systems. Ren et al. propose network motifs as an indicator of cascading outage risk \cite{RenPESGM17}. 
They show that cascading outages exhibit three phases as the load level increases, and the phases correspond to the decrease of the frequency of network motifs. 
The frequency of motifs reflects the connectivity of the power grid; hence, it can be a warning sign of the cascading outage risk. 
Other researchers have studied network motifs as an indicator of power grid robustness and reliability 
using techniques from network science \cite{poorSIP17,poorNAS19,abedijaberiICDMW18,poor21}. Specifically, they carry out attacks on the power grid by removing nodes according to some order, and monitor network motif properties such as concentration, z-score, and lifetime. 
Then they determine the robustness and reliability of the network based on the idea that a robust network tends to preserve longer its motif-based measurements. 

Researchers also study outage patterns using influence/interaction graphs \cite{qiPS15,juIEEEJournal17,hinesPS17,sunAccess19,zhouPS20,zhouPMAPS20}. An essential difference from this work is that they study propagation patterns of cascading outages, while this work aims at revealing spatial patterns of initial simultaneous outages for better contingency selection and risk estimation.
However, influence/interaction graphs generated from simulated cascades could be improved by using better contingency lists for the simulated cascades. Influence/interaction graphs generated from utility data can empirically account for the frequencies of initial outages \cite{zhouPS20}. It is also feasible \cite{zhouPMAPS20} to simulate an influence graph from assumed initial conditions, and this could use better contingency lists.  

The previous work on network motif applications in power systems uses the conventional definition of network motifs, which defines motifs as connected subgraphs in a network that occur significantly more frequently than in a random network \cite{MiloScience02}.
However, this definition is not well suited to contingency selection because the power network is not a random network; it is a particular network of known structure. Moreover, multiple contingencies can also be disconnected subgraphs. 
Therefore, in this paper, we newly define contingency motifs as connected or disconnected subgraphs that occur significantly more frequently than random subgraphs of the particular power network under consideration. 

\section{Multiple line initial outages frequently occur in contingency motifs}
\label{sec:dmotif}

We analyze 19 years of historical outage data recorded by Bonneville Power Administration (BPA) and publicly available at \cite{bpadata}. The first 14 years of data are used for analysis and the last 5 years of data are used for testing. The power transmission network is deduced from the outage data itself using the method in \cite{DobsonPS16}. 
The automatic line outages are grouped into cascades and then  generations according to the outage times\footnote{The grouping of line outages uses the method detailed in \cite{dobsonPS12}: Looking at the gaps in start time between successive line outages, if successive outages have a gap of one hour or more, then the outage after the gap starts a new cascade; if outages occur within the same minute, they are in the same generation of a cascade. 
More elaborate methods of grouping line outages to find initiating outages could also be applied.}.   
Then, multiple initial line outages are extracted from the first generations of cascades and represented as subgraphs of the power network. Some patterns are frequently recurrent, and we adapt the network motif concept to represent these frequently occurring patterns as {\sl contingency motifs}. 
%\looseness=-1  

This section first describes the statistics of random patterns of the power network and statistics of patterns observed in historical outages, and then it gives the definition and identification of contingency motifs. 

\subsection{Subgraphs and patterns in the power transmission network}
\begin{figure}[th!]
    \centering
    \includegraphics[width=0.8\columnwidth]{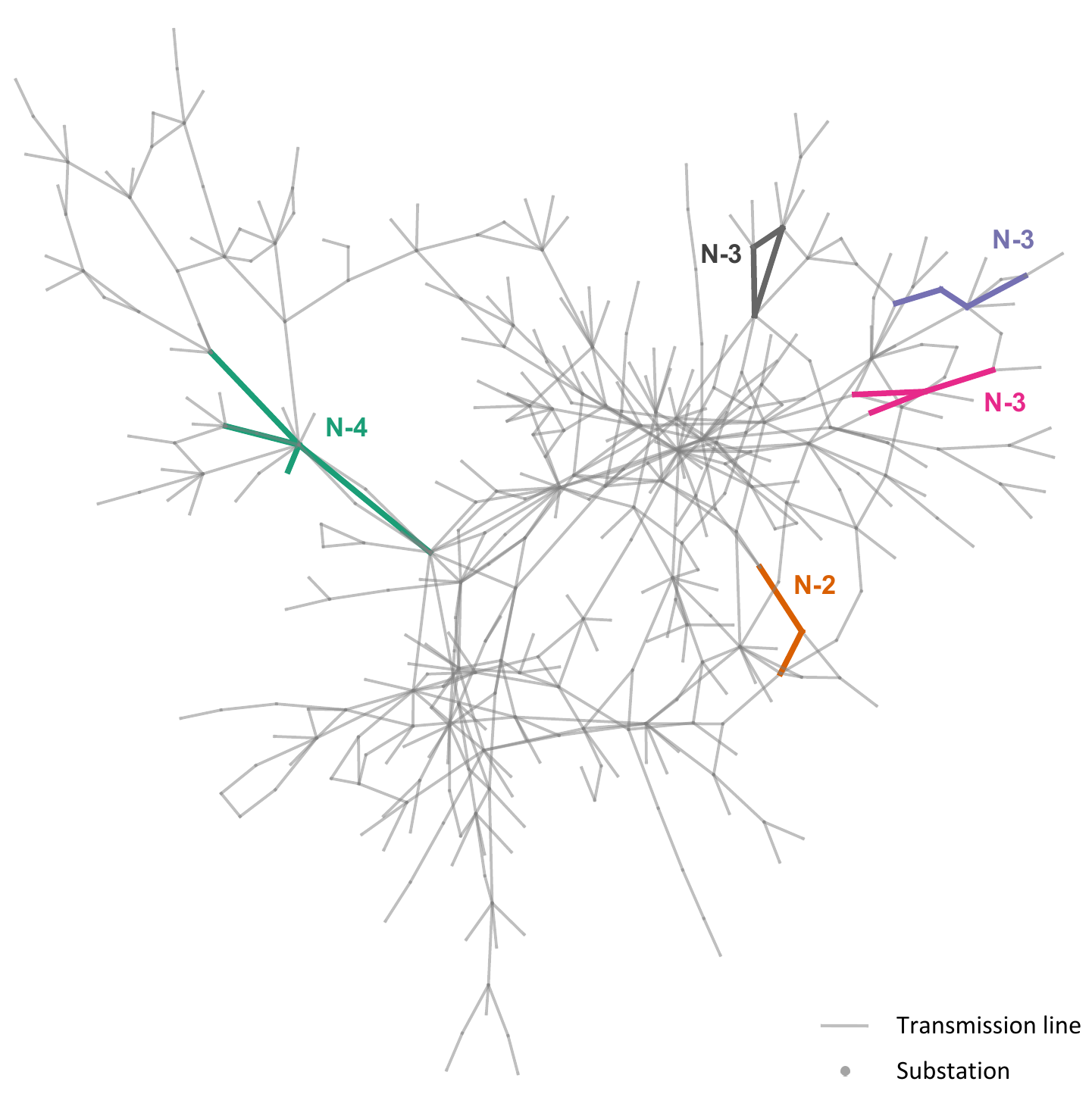}
    \caption{BPA power transmission network (528 lines) derived from the outage data \cite{DobsonPS16}. Highlighted subgraphs are five examples of multiple line outages. Layout is not geographic.}
    \label{fig:bpagraph}
\end{figure}

The BPA power transmission network  is shown in Figure~\ref{fig:bpagraph}. Substations correspond to network nodes, and transmission lines correspond to network edges. The power grid has multiple transmission lines between some substations, and they are represented by one line in this network.

\begin{figure}[!ht]
    \centering
    \includegraphics[width=0.4\columnwidth]{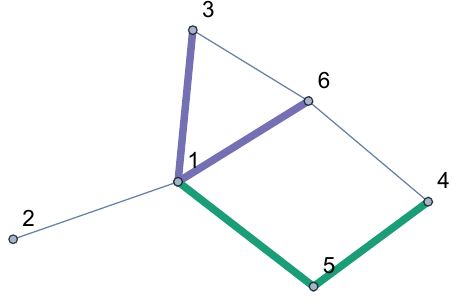}
    \caption{The two highlighted subgraphs are isomorphic 2-edge subgraphs.}
    \label{fig:egsubgraph}
\end{figure}

A $k$-edge subgraph is an edge-induced subgraph, which is a subset of edges of a graph together with nodes that are their endpoints. For example, $\{1-3,~1-6\}$ is a two-edge subgraph of the graph in Figure~\ref{fig:egsubgraph}.
When an N-k contingency occurs, we can imagine that the k outaged lines in the power network are highlighted, and we observe a subgraph. 
Thus, each N-k contingency corresponds to a subgraph.

Two subgraphs are isomorphic when there is a mapping between their nodes such that two nodes adjacent in one subgraph implies that the corresponding two nodes in the other subgraph are also adjacent. We say that two subgraphs are the same when their nodes and edges are exactly the same. For example, in Figure~\ref{fig:egsubgraph}, subgraphs $\{1-3,~1-6\}$ and $\{1-5,~4-5\}$ are isomorphic, but are not the same subgraph.  

A pattern is a set of isomorphic edge subgraphs. $S_{k,i}$ denotes a pattern that is a set of subgraphs $s_{k,i}$, where $k$ is the number of edges and $i$ is the pattern identifier. An exception is $S_{4,*}$, which denotes the set of 4-edge subgraphs that are not members of $S_{4,i}$ for $i = 1,2,3,4$. Table~\ref{tbl:nkpatterns} shows patterns of the BPA network and the number of distinct subgraphs in each pattern (the size $|S_{k,i}|$  of the pattern). As contingencies are always grouped according to the number of outaged components $k$, subgraphs are also grouped this way.
 \begin{table}[!ht]
	\caption{Probabilities of patterns in BPA data and random subgraphs}
	\label{tbl:nkpatterns}
	\centering
	\begin{tabular}{M{0.5cm}M{0.5cm}M{1.1cm}M{1.4cm}M{0.4cm}M{0.4cm}M{1.4cm}  }
		 & $S_{k,i}$& $|S_{k,i}|$&$P^{\rm uni}(S_{k,i}|k)$&$n_k$&$n_{k,i}$& $P(S_{k,i}|k)$\\
		\hline
		\includegraphics[width=5mm]{s21.pdf}  & $S_{2,1}$ & 2116 & 0.015 & 392 & 317 & 0.81 \\
        \includegraphics[width=5mm]{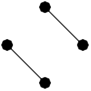}  & $S_{2,2}$ &  137\,012 & 0.98 & 392 & 75 & 0.19 \\
        &  &    &   &   &   &   \\
        \includegraphics[width=5mm]{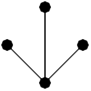}  & $S_{3,1}$ & 4653 & 0.0002 & 127 & 74 & 0.58 \\
        \includegraphics[width=5mm]{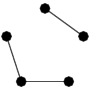}  & $S_{3,2}$ & 1\,083\,833 & 0.044 & 127 & 31 & 0.24\\
        \includegraphics[width=5mm]{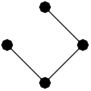}  & $S_{3,3}$ & 7519 & 0.0003 & 127 & 18 & 0.14 \\
        \includegraphics[width=5mm]{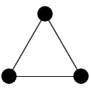}  & $S_{3,4}$ & 62 & $ 10^{-6}$ & 127 & 3 & 0.024 \\
        \includegraphics[width=5mm]{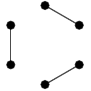}  & $S_{3,5}$ &23\,297\,709 & 0.96 & 127 & 1 & 0.0079 \\
        &   &   &   &   &   &   \\
        \includegraphics[width=5mm]{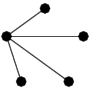}  & $S_{4,1}$ & 9799 & $ 10^{-6}$ & 23 & 9 & 0.39 \\
        \includegraphics[width=5mm]{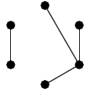}  & $S_{4,2}$ & 2\,354\,215 & $ 10^{-4}$ & 23 & 5 & 0.22 \\
        \includegraphics[width=5mm]{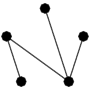}  & $S_{4,3}$ & 48\,581 & $ 10^{-5}$ & 23 & 5 & 0.22 \\
        \includegraphics[width=5mm]{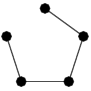}  & $S_{4,4}$ & 26\,028 & $ 10^{-6}$ & 23 & 2 & 0.087 \\
        others  & $S_{4,*}$  & 3\,199\,244\,477 & 0.99 & 23 & 2 & 0.087 \rule[-1.0ex]{-3pt}{0pt}\\
        \hline
        \multicolumn{7}{c}{\footnotesize $S_{4,*}$ is the set of 4-edge subgraphs that are not in $S_{4,i}$ for $i = 1,2,3,4$.}
	\end{tabular}
\end{table}

\subsection{Probability of patterns}
Let $P(S_{k,i}|k)$ be the probability that a pattern $S_{k,i}$ appears in contingency subgraphs given the number of lines $k$. Two methods are used to estimate $P(S_{k,i}|k)$: one based on the uniform assumption and the other based on outage data. 

\subsubsection{Uniform assumption}
Suppose the network has $N$ nodes.
If no other information is available, it is natural to assume that contingency subgraphs occur uniformly in all the $\binom{N}{k}$ possible $k$-edge subgraphs of the network. That is
\begin{align}
    p_{s_{k,i}}^{\rm uni} = \frac{1}{\binom{N}{k}} \label{equ:pski1}
\end{align}
where $p_{s_{k,i}}^{\rm uni}$ denotes the probability of a particular subgraph $s_{k,i}$ given $k$, and ``uni'' indicates uniform. We call this the uniform assumption.

$|S_{k,i}|$ is the number of subgraphs of the network in $S_{k,i}$. 
Then $P(S_{k,i}|k) $ under the uniform assumption is
\looseness = -1
\begin{align}
    P^{\rm uni}(S_{k,i}|k) =  \frac{|S_{k,i}|}{\binom{N}{k}} \label{equ:pki1}
\end{align}
as shown in the fourth column of Table~\ref{tbl:nkpatterns}.

\subsubsection{Empirical probability}
$P(S_{k,i}|k)$ estimated from the outage data is 
\begin{align}
    P(S_{k,i}|k) = \frac{n_{k,i}}{n_k} \label{equ:pki2}
\end{align}
where 
%``obs'' indicates the probability is estimated from outage data, 
$n_{k,i}$ is the number of contingency subgraphs $s_{k,i}$ appearing in the outage data, and $n_k$ is the number of $k$-edge contingency subgraphs in the network. Note $\sum_i n_{k,i} = n_k$. $P(S_{k,i}|k)$ is shown in the last column of Table~\ref{tbl:nkpatterns}.

\subsubsection{Discussion}
Table~\ref{tbl:nkpatterns} shows that
the probabilities from outage data differ greatly from the uniform assumption. 
For example, $P(S_{2,1}|2)$ is much greater than $P^{\rm uni}(S_{2,1}|2)$, and $P(S_{3,1}|3)$ is much greater than $P^{\rm uni}(S_{3,1}|3)$.
This implies that some patterns recur much more frequently than indicated by the uniform assumption. These frequently recurrent patterns %in contingency subgraphs are referred to as
are the contingency motifs discussed in the next subsection.  

\subsection{Contingency motif definition}
The conventional definition of a network motif \cite{MiloScience02} considers connected subgraphs with a specific number of nodes. For example, possible size-3 motifs in Figure~\ref{fig:egsubgraph} are subgraphs  $\{1-3,~1-6,~3-6\}$ and $\{1-5,~4-5\}$.  
Conventional motifs are detected by 
computing the frequency of each pattern and comparing it with the frequency of the same pattern in random networks with the same global property, such as degree distribution, as the original network \cite{MiloScience02, WernickeFanmod06}. 
However, the power network is a particular, non-random network, contingency subgraphs can be disconnected subgraphs, and contingencies are grouped according to the number of lines, not the number of nodes. 
Therefore, the conventional definition of the network motif cannot be directly applied, and we define contingency motifs as follows.

Instead of comparing the frequency of a pattern in outage data to that in a random network, we compare the frequency of the pattern to that in subgraphs sampled randomly from the particular power network under consideration.
We define a $k$-edge contingency motif in a power network as a $k$-edge pattern whose probability of occurrence is significantly greater than that when all $k$-edge subgraphs in the power network are assumed to have the same probability of occurrence. That is,
\begin{align}
    P(S_{k,i}|k) > a P^{\rm uni}(S_{k,i}|k)
    \label{significant}
\end{align}
where $a \geq 1$ is large enough to indicate a significant difference. We choose $a = 10$ in this paper.
For example, to determine 3-edge motifs, we estimate the probability of $S_{3,i}$ for all $i$ from outage data, and then compute the probability of $S_{3,i}$ in the network under the uniform assumption. If the probability of a pattern in outage data is significantly statistically greater than that under the uniform assumption according to (\ref{significant}), the pattern is a contingency motif.

\subsection{Contingency motifs in the power network}
\looseness=-1
To detect a contingency motif from  data, we compare the probability of the contingency pattern observed in outage data and the probability of that pattern under the uniform assumption. This problem can be formulated as a hypothesis test:
\begin{align}
    H_0: P(S_{k,i}|k) \leq  10 P^{\rm uni}(S_{k,i}|k) \notag \\\text{versus} \qquad H_1: P(S_{k,i}|k) > 10 P^{\rm uni}(S_{k,i}|k)\notag
\end{align}

\looseness=-1
We use frequentist and Bayesian methods to do the hypothesis test as detailed in Appendix \ref{app1}, and both tests identify the same motifs: $S_{2,1}$ is a 2-edge contingency motif, $S_{3,1}$, $S_{3,3}$, $S_{3,4}$ are 3-edge contingency motifs, and $S_{4,1}$, $S_{4,2}$, $S_{4,3}$, $S_{4,4}$ are 4-edge contingency motifs. 
The test results are shown in Table~\ref{tbl:nkmotif}, including 
the p-value in the frequentist hypothesis test and the posterior probability $P(H_0 | n_{k,i})$ in the Bayesian hypothesis test. 

 \begin{table}[!ht]
	\caption{Contingency motifs of the BPA data}
	\label{tbl:nkmotif}
	\centering
	\begin{tabular}{M{0.8cm}M{0.8cm}M{1.4cm}M{1.4cm}M{1.4cm}  }
		 & $S_{k,i}$ & motif & p-value & $P(H_0 | n_{k,i})$\\\hline
		\includegraphics[width=5mm]{s21.pdf}  & $\bm S_{2,1}$ & true & 0. & 0.  \\
        \includegraphics[width=5mm]{s22.pdf}  & $S_{2,2}$ & false & 1. & 1.  \\
        &  &    &   &     \\
        \includegraphics[width=5mm]{s31.pdf}  & $\bm S_{3,1}$& true & 0. & 0.  \\
        \includegraphics[width=5mm]{s32.pdf}  & $S_{3,2}$& false & 1. & 0.99  \\
        \includegraphics[width=5mm]{s33.pdf}  & $\bm S_{3,3}$  & true& 0. & 0. \\
        \includegraphics[width=5mm]{s34.pdf}  & $\bm S_{3,4}$& true & 0. & 0.  \\
        \includegraphics[width=5mm]{s35.pdf}  & $S_{3,5}$ & false& 1. & 1.  \\
        &   &   &   &    \\
        \includegraphics[width=5mm]{s41.pdf}  & $\bm S_{4,1}$ & true& 0. & 0.  \\
        \includegraphics[width=5mm]{s42.pdf}  & $\bm S_{4,2}$ & true& 0. & $10^{-8}$  \\
        \includegraphics[width=5mm]{s43.pdf}  & $\bm S_{4,3}$ & true& 0. & 0.  \\
        \includegraphics[width=5mm]{s44.pdf}  & $\bm S_{4,4}$ & true & 0. & $10^{-9}$  \\
        others  & $S_{4,*}$ & false & 1. & 1.  \\
	\end{tabular}
\end{table}

\subsection{Discussion}

We only consider transmission line outages.
%, and node outages (such as generator and transformer outages) are not directly considered. However, line outages usually accompany node outages, and they often have equivalent effects on the power flow model. 
In terms of physical elements in power systems, multiple contingencies involve primary devices (generators, lines, transformers, compensators, circuit breakers, bus-bar sections) and secondary devices (protections and telecommunication equipment). Outages of these devices can result in both single and multiple contingencies of transmission lines. 
NERC standard \mbox{TPL-001-4} describes seven categories of contingencies related to various devices \cite{nercstandard}, and they can be further grouped into four types: N-1, {N-1-1}, N-2, and N-k for k$>$2. For example, category P3 are single-phase short circuit to ground faults of a bus-bar section; if the bus-bar section connects k lines, then an N-k line contingency occurs.

Multiple line outages can be divided into dependent and independent contingencies. Dependent outages are closely related to bus configurations and protective relays. It needs a lot of effort to build a detailed power system model including relays \cite{dobsonPMAPS18}. Scheduled maintenance and forced outages change the topology of the power network, and hidden failures in the protective relay system are inevitable. There are also common-mode multiple contingencies\footnote{A common-mode contingency is a multiple contingency caused by a single event where outages are not consequences of each other. For example, a single lightning stroke can cause two line outages on a common tower.} that are caused by extreme weather or other external factors \cite{papicGM12}. 

Two-edge stars  \includegraphics[height=9pt]{s21.pdf} in $S_{2,1}$ could be two transmission lines connected to the same substation faulted simultaneously by coincidence, common-mode contingencies of two lines, a circuit breaker or a tie break stuck in a breaker and a half substation, primary protection fails and zone 2 protection is activated, a fault in a bus-bar section connecting two transmission lines, or a hidden relay system failure, etc..
In the first cause, the two line outages are independent because one line outage does not cause the other line outage; while for the rest of the causes, the two line outages are dependent on physical or engineered structure. 
Thus, $S_{2,1}$ as a motif usually reflects some inherent dependence of two lines. 

The causes for three-edge and four-edge stars could include faults of transmission lines connected on the same bus-bar section, faults of bus-bar sections, transformers outages, breaker stuck, etc. $S_{3,3}$ is composed of three lines in a row. A possible cause is that these three lines are in a protection control group. $S_{3,4}$ is a triangle, which is a special local structure in the power network that is limited in number. 
%N-4 are rare and the three connected patterns turn out to be motifs. We speculate that the causes could be substation outages or backup protection systems are activated. 

\looseness=-1
The precise physical or engineering dependencies causing a specific motif are not clear without detailed knowledge of a system, but the existence of the motif underlines the importance of studying multiple contingency mechanisms in detail.

\section{Probabilities of multiple line outages}
\label{sec:prob}

\subsection{A probabilistic model for multiple line outages}
As multiple line outages show different patterns and some are contingency motifs, we partition the whole contingency space accordingly. Moreover, some patterns are disconnected subgraphs, and they have different network diameters{\footnote{The diameter of a subgraph is the largest network distance between any two lines, and the network distance of two lines in a subgraph is the minimum number of nodes of a network path connecting the two lines \cite{DobsonPS16,estradaCICS96}.}}. As the network diameter follows a Zipf distribution as discovered in \cite{zhouThesis22}, we further partition disconnected patterns according to their diameter. The Zipf distribution has a heavy tail, implying that multiple line outages containing far-away lines do occur.  

\begin{figure}[!ht]
    \centering
    \includegraphics[width=0.7\columnwidth]{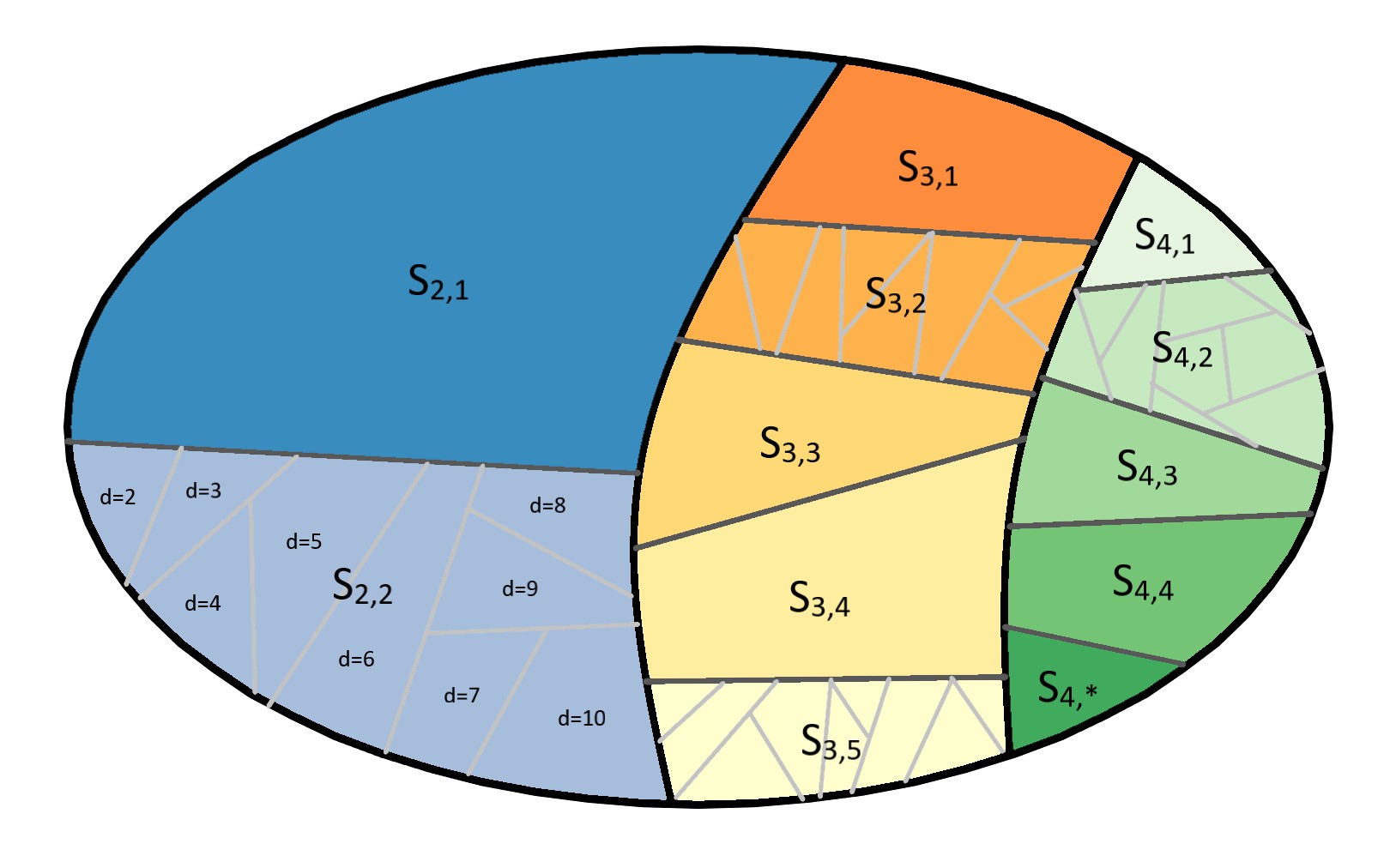}
    \caption{Partition of the contingency subgraph space. Each cell represents a pattern $S_{k,i}$ 
    with a specific diameter $d$. Multiple contingencies $s_{k,i}$ in each cell are assumed to have equal probabilities.} 
    \label{fig:illustration}
\end{figure}

The partition is illustrated in Figure~\ref{fig:illustration}. The ellipse represents the space of contingency subgraphs, including N-2, N-3, and N-4. According to the different patterns, N-k contingencies are further divided into groups $S_{k,i}$. Furthermore, disconnected $S_{k,i}$ are divided into subgroups according to their diameters. Each cell represents a $S_{k,i}$ with a specific diameter $d$. 
A key assumption is that multiple contingencies $s_{k,i}$ in each cell have equal probabilities. 

We build a probabilistic model to estimate the probability of the multiple line outage $s_{k,i}$ with $k$ lines  based on the statistics of outage data. That is,
\begin{align}
     P &\left(s_{k,i}\right)  = P\left(k, S_{k,i}, d, s_{k,i}\right)  \notag \\
    %= & P\left( k\right) P\left(S_{k,i}, d, s_{k,i} | k\right) \notag \\
    %= & P\left( k\right) P\left( S_{k,i} | k \right) P\left(s_{k,i}, d | k, S_{k,i} \right) \notag \\
    & =  P\left( k\right) P\left( S_{k,i} | k \right) P\left( d | k, S_{k,i} \right) P\left( s_{k,i} |k, S_{k,i}, d \right) \notag \\
    & = P\left( k\right) P\left( S_{k,i} | k \right) P\left( d | S_{k,i} \right) P\left( s_{k,i} | S_{k,i}, d \right) 
     \label{equ:ps1}
\end{align}
where $P\left( k\right)$ is the probability of $k$ line outages, 
%$P\left( S_{k,i} | k\right)$ is the probability of pattern $S_{k,i}$ given $k$ line outages, 
$P\left( d | S_{k,i}\right)$ is the probability that pattern $S_{k,i}$ has diameter $d$, and $P\left( s_{k,i} | d, S_{k,i}\right)$ is the probability of a specific multiple contingency given its pattern and diameter.  

\subsubsection{Probability of the number of line outages}
It is natural to estimate the probability $P\left( k\right)$ by
\begin{align}
    P(k) = \frac{n_{k}}
%    {\sum_{k=2}^{4} n_{k}}
    {n_2+n_3+n_4}
    , \qquad k = 2,3,4 \label{equ:k}
\end{align}
The distribution of $k$ for the BPA data is $P(k=2) = 0.72,~P(k=3) = 0.24,~P(k=4) = 0.04$. 
N-k contingencies for k$>$4 are not considered because of their very rare occurrence. 

\subsubsection{Probability of a pattern given $k$ line outages}
$P\left( S_{k,i} | k\right)$ is estimated during the detection of contingency motifs and is shown in Table~\ref{tbl:pmotif}. 
\begin{table}[!ht]
    \centering
    \caption{Distribution of patterns $P\left( S_{k,i} | k\right)$ for BPA data.}
    \label{tbl:pmotif}
    \begin{tabular}{cccccc}
        $S_{k,i}$ & $S_{2,1}$  & $S_{2,2}$  & & & \\\hline
        $P(S_{k,i} | k)$ & 0.809 & 0.191 &  & &  \\
        $S_{k,i}$ &$S_{3,1}$  & $S_{3,2}$ & $S_{3,3}$  & $S_{3,4}$ & $S_{3,5}$ \\\hline
        $P(S_{k,i} | k)$ &  0.583 & 0.244 & 0.141 & 0.024 & 0.008 \\
        $S_{k,i}$ & $S_{4,1}$ & $S_{4,2}$ & $S_{4,3}$ & $S_{4,4}$ & $S_{4,*}$\\\hline
        $P(S_{k,i} | k)$  & 0.391 & 0.217 & 0.217 & 0.087 & 0.087
    \end{tabular}
\end{table}

\looseness=-1
\subsubsection{Probability of the contingency diameter given its pattern}
The connected contingency subgraphs have patterns $ S_{2,1}$, $ S_{3,1} $, $ S_{3,3} $, $ S_{3,4} $, $ S_{4,1} $, $ S_{4,3} $, $ S_{4,4} $. For the connected contingency subgraphs, the diameter is constant or very nearly constant\footnote{$s_{2,1}$, $s_{3,1}$, $s_{3,4}$,  $s_{4,1}$ have diameter 1 and $s_{3,3}$, $s_{4,3}$  have diameter 2. Although the diameter of $s_{4,4}$ is 2 or 3, 98\% of $s_{4,4}$ have a diameter of 3 in the BPA network.}. Therefore, we take the diameter distribution of the connected contingency subgraphs to have probability 1 at a constant diameter. 

The disconnected contingency subgraphs have patterns $S_{2,2}$, $S_{3,2}$, $S_{3,5}$, $S_{4,2}$.
For all the disconnected contingencies combined together,
we empirically estimate from the outage data the distribution of diameter $P(d|\mbox{disconnected})$.
The disconnected contingencies are all combined together to calculate the single distribution $P(d|\mbox{disconnected})$ because of the limited outage data for these subgraphs.  $P(d|\mbox{disconnected})$ is the number of disconnected subgraphs with diameter $d$ divided by the total number of disconnected subgraphs.

$S_{k,*}$ has both connected and disconnected subgraphs, but a small probability. Therefore 
%Since $S_{k,*}$ has a small probability but contains many different patterns, it is convenient to assume subgraphs in $S_{k,*}$ are uniformly distributed. 
we set the diameter distribution to have probability 1 as for the connected subgraphs. However, we can only determine the value of $d$ when a specific $s_{k,*}$ is given to make it a valid probability distribution, and if $s_{k,*}$ is given, $d$ equals the diameter of this specific $s_{k,*}$.

In summary, the diameter distribution conditional on pattern $S_{k,i}$ is estimated by
\begin{align}
     P\left(d | S_{k,i}\right) 
    = 
    \begin{cases}
    1 & S_{k,i} \mbox{ connected} \\
   P(d|\mbox{disconnected})& S_{k,i} \mbox{ disconnected}\\
    1 & S_{k,*}\\
    \end{cases} \label{equ:dists1}
\end{align}
Note that (\ref{equ:dists1}) does not explicitly express the support (all possible values) of $d$  for the different $S_{k,i}$ because it is obvious. For example, $S_{2,1}$ has $d \in \{1\}$; 
%$S_{k,*}$ has an undetermined diameter, and for a specific $s_{k,*}$, we always take probability 1 whatever the diameter is.

\subsubsection{Probability of a contingency given its pattern and diameter}
Finally, assume that subgraphs of a pattern $S_{k,i}$ with diameter $d$ are uniformly distributed. That is, $P\left( s_{k,i} | S_{k,i}, d\right)$ is the discrete uniform distribution
\begin{align}
    P\left( s_{k,i} | S_{k,i}, d)\right) = \frac{1}{|S^d_{k,i}|} \label{equ:unifd}
\end{align}
$|S^d_{k,i}|$ denotes the number of subgraphs in $S_{k,i}$ with diameter $d$. $|S^d_{k,i}|$ is approximated by uniformly sampling a large number of $s_{k,i}$ and computing their diameters, as shown in Table~\ref{tbl:diam}.
\begin{table}[!ht]
    \centering
    \caption{Number of distinct subgraphs with different diameters in $S_{k,i}$}
    \label{tbl:diam}
    \begin{tabular}{ccccc}
    $d$ & $|S^d_{2,2}|$ &$|S^d_{3,2}|$ & $|S^d_{3,5}|$ & $|S^d_{4,2}|$\\\hline
    2 & 6592& 47035&10129 & 110725\\
    3 & 14330 &133813& 333798& 232684\\
    4 &22607&  203860& 1442182 &414143\\
    5&27069&231926&3395950 &522119\\
    6& 25360& 201685& 5041582 &475201\\
    7& 18777&133339&5069585&318717\\
    8& 11653&73988&3781098&160323\\
    9& 6283&36730&2327380&75294\\
    10&2897& 15178& 1213277&34563
    \end{tabular}
\end{table}

\subsection{Probabilities of multiple line outages}
Given a specific multiple contingency $s_{k,i}$, 
we can estimate its probability with (\ref{equ:ps1}) by substituting values in Table~\ref{tbl:pmotif}, and computing probabilities with (\ref{equ:k}), (\ref{equ:dists1}),  (\ref{equ:unifd}).
Table~\ref{tbl:ppattern} shows the probability of any contingency $s_{k,i}$ in a pattern with diameter $d$. $S_{4,*}$ is not included because it has a great number of distinct $s_{4,*}$, and thus each $s_{4,*}$ has an extremely small probability. Table~\ref{tbl:ppattern} confirms that motifs have higher probabilities than other patterns.   

\begin{table}[!ht]
    \centering
    \caption{Probability of outages with different patterns and diameters}
    \label{tbl:ppattern}
    \resizebox{\columnwidth}{!}{ 
    \begin{tabular}{cccccccccccc}
        d & $s_{2,1}$& $s_{2,2}$ & $s_{3,1}$ & $s_{3,2}$ & $s_{3,3}$ & $s_{3,4}$ &$s_{3,5}$ & $s_{4,1}$ & $s_{4,2}$ & $s_{4,3}$ & $s_{4,4}$ \\ \hline
        1& 3E-4 &$ 0. $& 3E-5 &$ 0. $&$ 0. $& 9E-5 &$ 0. $& 3E-6 &$ 0.$ &$ 0. $&$ 0. $\\
        2&$0. $& 1E{-5} &$ 0. $& 7E{-7} & 2E{-6} & 0. & 1E{-9} & 0. & 4E{-8} & 4E{-8} & 0. \\
        3&0. & 2E{-6} & 0. & 9E{-8} & 0. & 0. & 1E{-9} & 0. & 8E{-9} & 0. & 4E{-8}\\
        4&0. &6E{-7} & 0. & 3E{-8} & 0. & 0. & 1E{-10} & 0. & 2E{-9} & 0. & 0. \\
        5&0. & 3E{-7} & 0. & 2E{-8} & 0. & 0. & 3E{-11} & 0. & 1E{-9} & 0. & 0. \\
        6& 0. &2E{-7} & 0. & 1E{-8} & 0. & 0. & 1E{-11} & 0. & 7E{-10} & 0. & 0. \\
        7&0. & 2E{-7} & 0. & 1E{-8} & 0. & 0. & 1E{-11} & 0. & 7E{-10} & 0. & 0. \\
        8& 0. & 2E{-7} & 0. & 2E{-8} & 0. & 0. & 1E{-11} & 0. & 1E{-9} & 0. & 0. \\
        9& 0. & 3E{-7} & 0. & 2E{-8} & 0. & 0. & 1E{-11} & 0. & 2E{-9} & 0. & 0. \\
        10& 0. & 3E{-7} & 0. & 2E{-8} & 0. & 0. & 1E{-11} & 0. & 2E{-9} & 0. & 0. 
    \end{tabular}}
\end{table}

\section{Forming a contingency list}
\label{sec:sampleA}
A contingency list is a sample of contingencies to assess the risk of cascading outages and other system violations such as line flows and voltages exceeding limits. 
The risk of cascading outages is often defined as the expected value of the impact \cite{camsPS12}. 
Three factors are considered in estimating the risk: (1) the probability of a contingency; (2) the probability distribution of cascading outage sizes, whose uncertainty also comes from pre-contingency system states, model parameters, and how the cascade evolves; (3) the size and impact of the blackout. 
The cascade size and impact are usually estimated through power system simulation. The risk $R(s)$ of contingency $s$ with impact $c(s)$ is 
$R(s) = P(s)E(c(s))$, where $P(s)$ is the probability of contingency $s$ and
$E(c(s))$
is the expectation of the cascade impact\footnote{For a deterministic simulation,  $E(c(s))$ reduces to $c(s)$.}, which can be estimated by Monte Carlo simulation given the initial contingency sample $s$. The overall system risk is then the average of the risk of individual contingencies. 
A contingency list that efficiently samples from a large fraction of the probable contingencies is fundamental to this risk calculation and is the subject of this paper.  Therefore, this section forms a contingency list from the contingency motifs for the BPA data. 

\subsection{Straightforward sampling of  contingencies}
\begin{figure}[!ht]
    \centering
    \includegraphics[width=0.6\columnwidth]{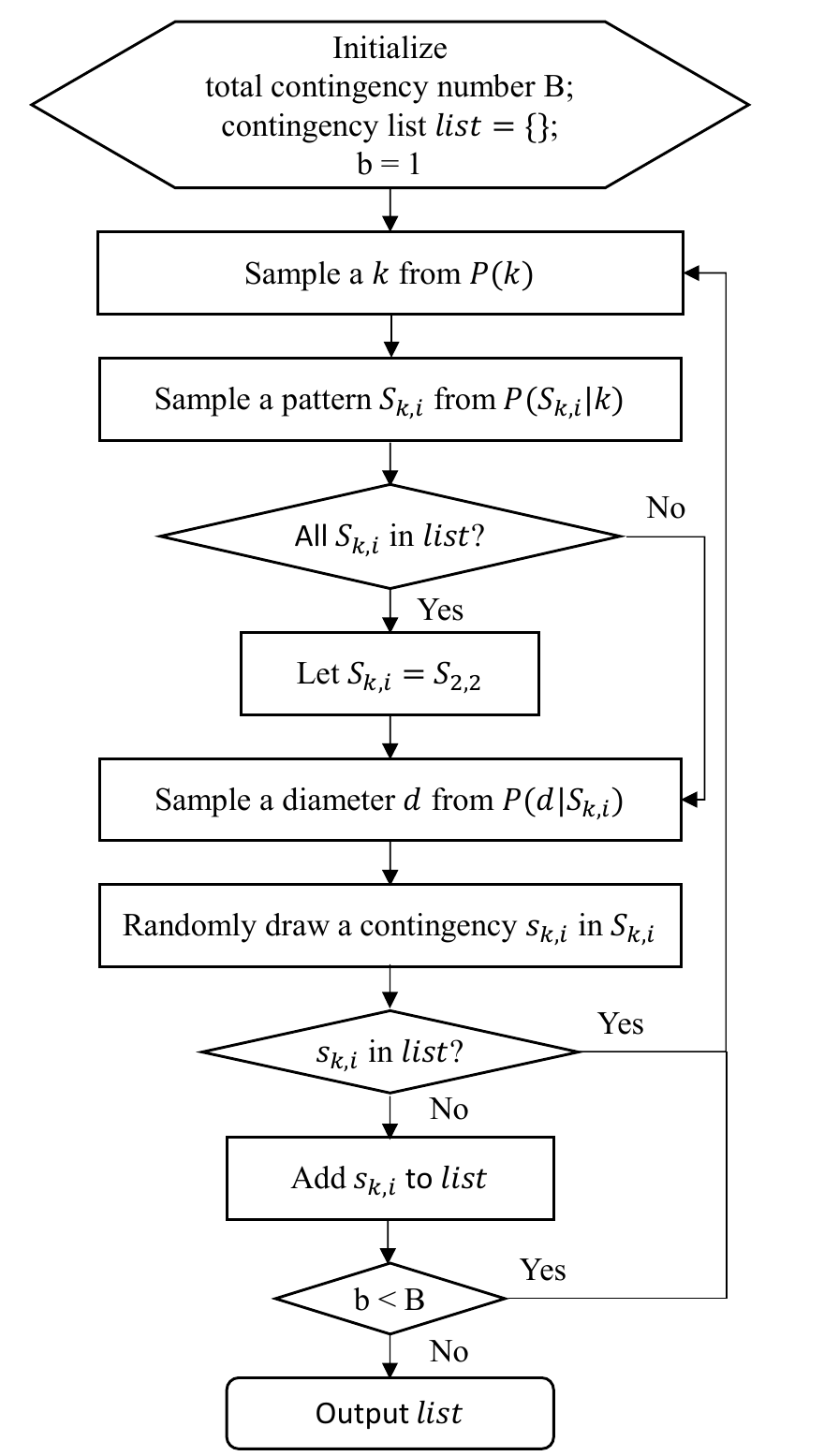}
    \caption{Flowchart of the straightforward sampling.}
    \label{fig:flowchart}
\end{figure}

The probabilistic model of (\ref{equ:ps1}) implies a straightforward sampling scheme for multiple line outages with four steps: (1) sample $k$ according to $P(k)$; (2) sample $S_{k,i}$ according to $P(S_{k,i}|k)$; (3) sample diameter $d$ according to $P(d|S_{k,i})$; (4) sample a $s_{k,i}$ uniformly from all subgraphs in $S_{k,i}$ with diameter $d$. Figure~\ref{fig:flowchart} shows the flowchart of sampling a contingency list including $B$ distinct contingencies. 
%\begin{figure}[!ht]
%    \centering
%    \includegraphics[width=0.8\columnwidth]{Figures/egSampleski1.pdf}
%    \caption{Sampling (a) a $s_{2,2}$ motif and (b) a $s_{3,2}$ motif.}
%    \label{fig:egdrawski}
%\end{figure}

\begin{figure}[!ht]
     \centering
     \begin{subfigure}[b]{0.49\columnwidth}
         \centering
         \includegraphics[width=\textwidth]{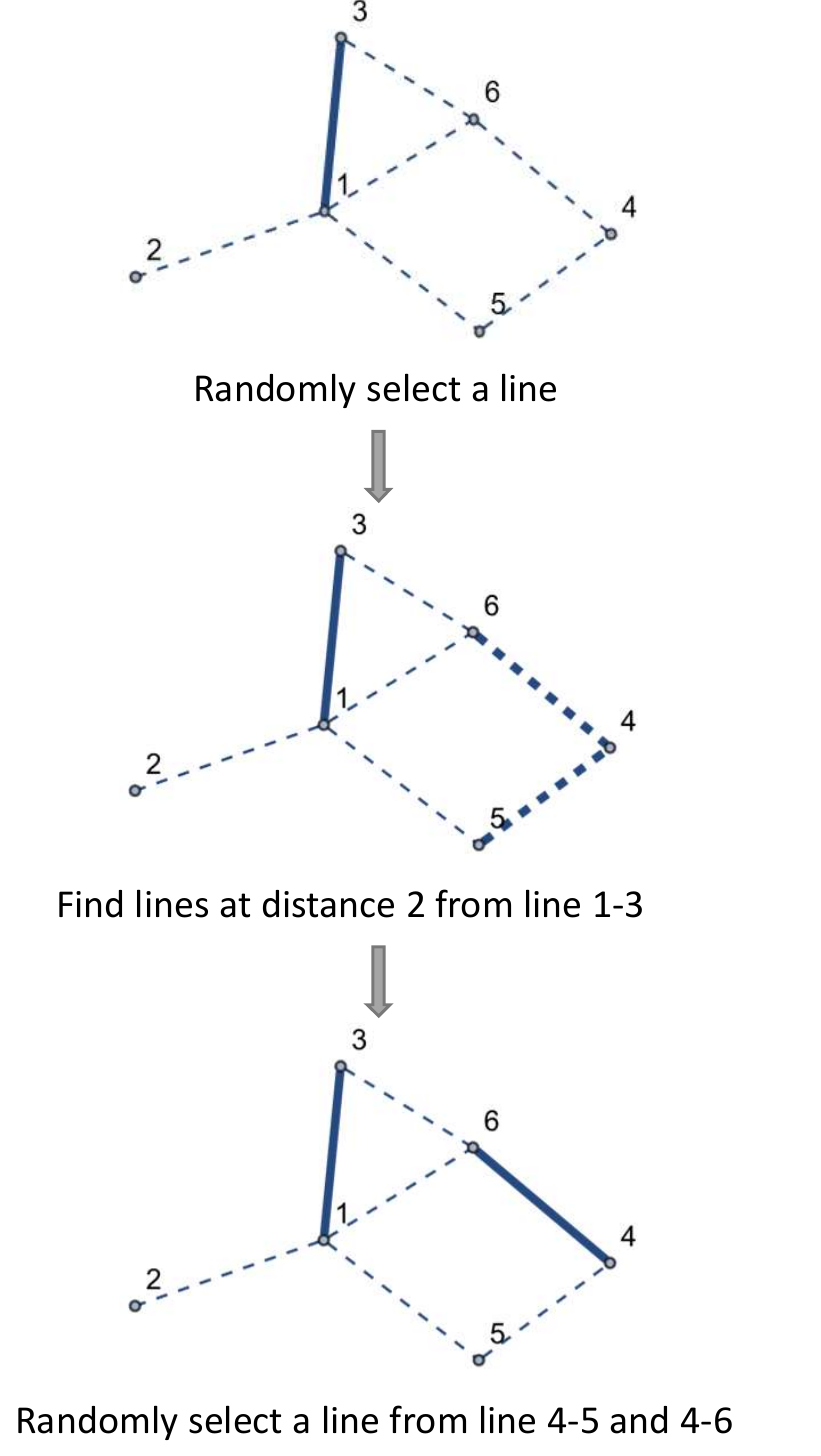}
         \caption{$s_{2,2}$}
         \label{fig:egdrawski_a}
     \end{subfigure} 
     \begin{subfigure}[b]{0.49\columnwidth}
         \centering
         \includegraphics[width=\textwidth]{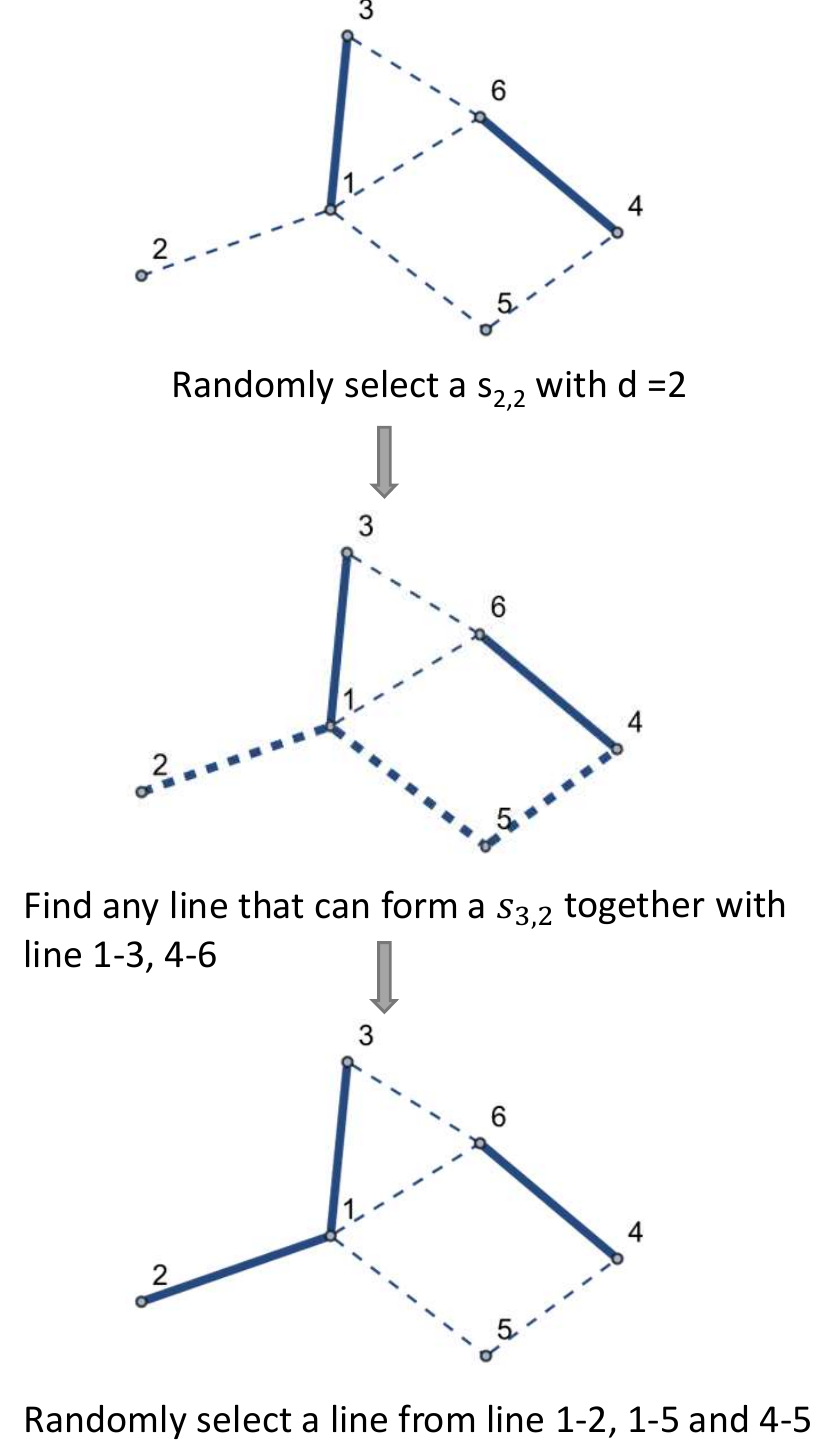}
         \caption{$s_{3,2}$}
         \label{fig:egdrawski_b}
     \end{subfigure}
        \caption{Sampling (a) a $s_{2,2}$ motif and (b) a $s_{3,2}$ motif.}
        \label{fig:egdrawski}
\end{figure}

\looseness=-1
The first three steps are straightforward as the corresponding random variables have a small number of discrete values. The fourth step is tricky because there is no effective way to find all subgraphs $s_{k,i}$ with pattern $S_{k,i}$ and diameter $d$, and randomly draw one of these subgraphs. For example, it is difficult to describe the 1\,083\,833 subgraphs in $S_{3,2}$.
Instead, we sample a $s_{k,i}$ by drawing lines sequentially. For N-2, first draw a line randomly; then find all lines that are at distance $d$ from the first line; finally, randomly draw a second line so that the first line and the second line form an N-2. Figure~\ref{fig:egdrawski_a} illustrates the steps of sampling a $s_{2,2}$ with diameter 2 using the small system in Figure \ref{fig:egsubgraph}. 
For N-3, draw the first line randomly and draw the second line that is at distance $d$ from the first line, as we do for N-2; then randomly draw the third line from lines that have a distance not greater than $d$ from either the first or the second line and form the desired pattern together with the previous two lines. Figure~\ref{fig:egdrawski_b} illustrates the steps of sampling a $s_{3,2}$ with diameter 2. 
For connected N-4, the distance is fixed. We first sample an N-3 that is a subgraph of the desired pattern, then sample the last line randomly from lines that can form the desired N-4. For $s_{4,2}$, we first draw a 3-edge star $s_{3,1}$ and then draw a line that is maximum distance $d$ from any of the three lines in the star. 
For $s_{4,*}$, we randomly sample 4 lines; if they do not form $s_{4,*}$, we sample again until they form a $s_{4,*}$. We will get a $s_{4,*}$ with only a few trials because its probability under the uniform assumption is high as shown in Table~\ref{tbl:nkpatterns}.

Using the sampling scheme, we draw 10\,000 N-2, N-3, and N-4. It is possible that we sample a contingency that is already sampled. In this case, we discard this contingency so that there is no repetition in the samples. This is actually sampling without replacement, which is less variable than sampling with replacement \cite{cochran77}. 
The three most likely contingencies at the top of the contingency list are $\{\{ 348- 365, 348 - 385\}, \{ 342- 378, 350 - 378\}, \{ 340- 353, 340 - 354\}\}$. They are in $S_{2,1}$ and have the same probability 0.0003. 

\begin{figure}[!ht]
    \centering
    \includegraphics[width=\columnwidth]{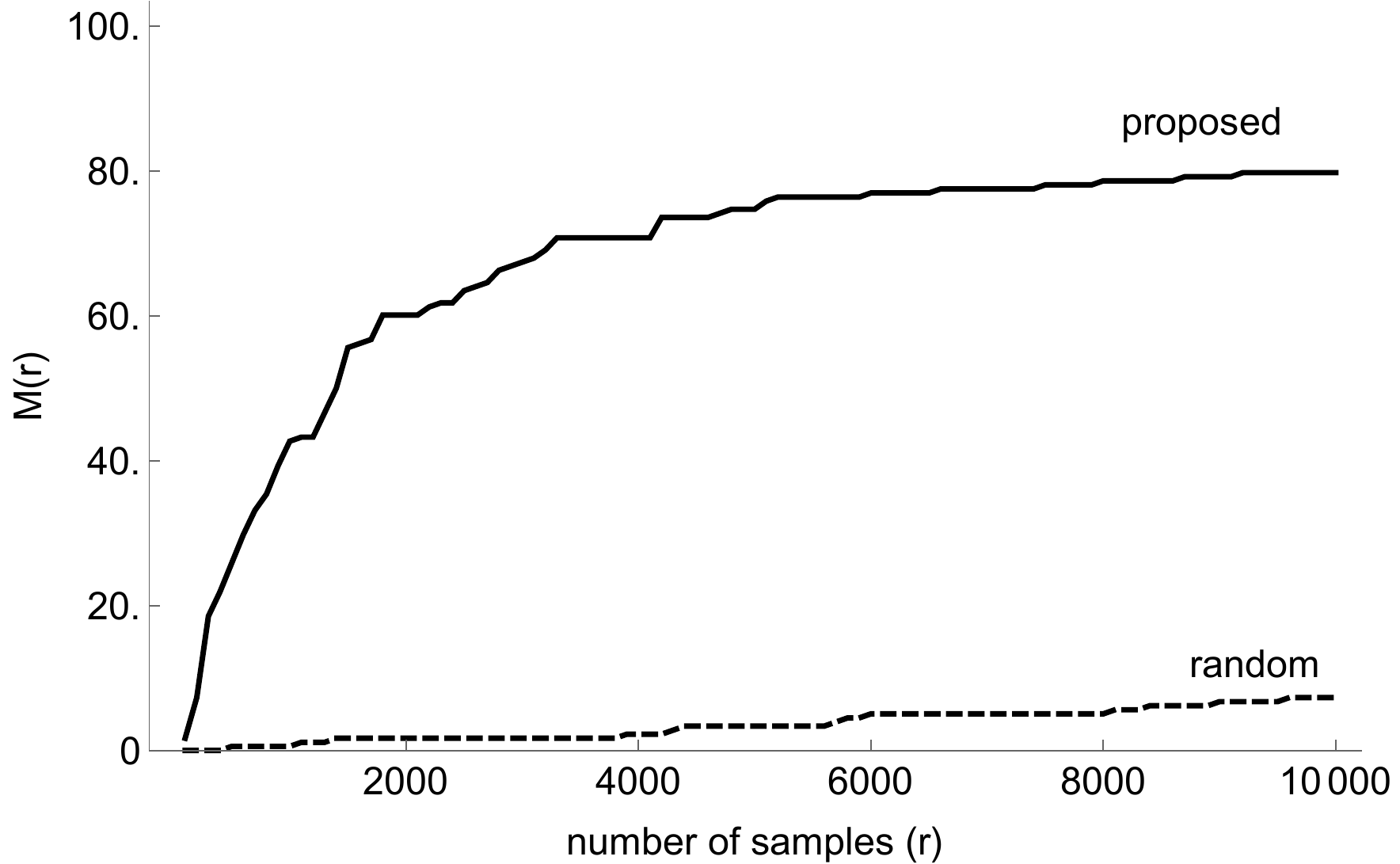}
    \caption{$M(r)$ for the straightforward sampling and the random sampling. The curves do not start at 0 because we compute $M(r)$ for $r=100,200,300,...$}
    \label{fig:cmpscheme}
\end{figure}
\begin{comment}
\begin{figure}[!ht]
    \centering
    \includegraphics[width=0.8\columnwidth]{Figures/cmpslope2.pdf}
    \caption{The slopes of $M(r)$ at $r=100,200,300,...$}
    \label{fig:cmpslope}
\end{figure}    
\end{comment}

\looseness=-1
Let $M(r)$ be the percentage of outages in the test data (last 5 years of outage data) that are covered by a contingency list with $r$ contingencies. To evaluate the performance of the straightforward sampling scheme, we compute $M(r)$ of the proposed contingency list and compare it with the same size list produced by a random scheme that treats all N-k as equally likely. That is, the random scheme samples an N-k by drawing a $k$ according to $P(k)$ and then drawing $k$ lines randomly from all lines. 

Figure~\ref{fig:cmpscheme} shows how $M(r)$ increases as $r$ increases. The straightforward sampling is much more efficient than the random sampling. Since we are using a sampling method, we draw ten lists with size $r$ to estimate the mean and standard deviation of $M(r)$. For the straightforward sampling, the average $M(10\,000)$ is 82\% with standard deviation 2\%; while for the random sampling, the average  $M(10\,000)$ is only 10\% with standard deviation 3\%.

The straightforward sampling scheme is designed so that outages with high probabilities are more likely to be drawn at an early stage. As a result, the solid curve in Figure~\ref{fig:cmpscheme} has a high slope at the beginning and then the curve flattens near the kink at $r = 3000$. 
When the slope of the straightforward sampling is higher than the random sampling, we are in a region where increments of effort in further sampling perform better than the random sampling. The kink is one possible indicator of stopping further sampling. Thus, we can use the first 3000 multiple contingencies to form a contingency list for detailed analysis. They cover about 70\% of the outages in the test data; in contrast, the first 3000 random contingencies only cover about 4\%.

The straightforwardly sampled contingencies %according to the probabilistic model 
have high coverage of outages in test data. It shows that 
%the distribution of multiple contingencies for the BPA data does not change significantly over a long period of time (19 years), and 
the contingency motifs capture the spatial statistics of multiple line outages.  

\subsection{Stratified sampling of contingencies}
The straightforward sampling scheme can be improved to stratified sampling, using motifs as strata. It is easy and flexible to implement and leads to more precise risk estimates than straightforward sampling.    

Three contingency motifs ($S_{2,1}$ \includegraphics[height=9pt]{s21.pdf}, $S_{3,1}$ \includegraphics[height=9pt]{s31.pdf}, and $S_{3,4}$ \includegraphics[height=9pt]{s34.pdf}) in the sampled contingencies in Section \ref{sec:sampleA}-A account for 78\% of the probability of multiple outages in the test data. 
As shown in Table~\ref{tbl:ppattern}, any individual $s_{2,1}$, $s_{3,1}$, or $s_{3,4}$ also has a higher probability than others. Another reason that we only consider these three motifs is that other motifs have a large number of distinct subgraphs (see Table \ref{tbl:nkpatterns}) and these three motifs can explain most of the probability. 
Therefore, we choose each of the motifs $S_{2,1}$, $S_{3,1}$, or $S_{3,4}$ as a stratum and all other patterns as a single stratum. Accordingly, (\ref{equ:ps1}) is rewritten as 
\begin{align}
     P\left(s_{k,i}\right) = &  P\left(s_{k,i}\right | S_{2,1}) P(S_{2,1}) + P\left(s_{k,i}\right | S_{3,1}) P(S_{3,1}) + \notag\\ 
     & \quad P\left(s_{k,i}\right | S_{3,4}) P(S_{3,4}) + 
     P\left(s_{k,i}\right | S_{*}) P(S_{*})
     \label{conditionalP}
\end{align}  
where $S_*$ represents any pattern that is not $S_{2,1}$, $S_{3,1}$, or $S_{3,4}$. Note that only one of the four terms on the right-hand side of (\ref{conditionalP}) is not zero for a specific $s_{k,i}$, so $P(S_{k,i}) = P(S_{k,i} | k) P(k)$, and $P(S_*)= 1-P(S_{2,1})-P(S_{3,1})-P(S_{3,4})$.
Directly calculating with the probability estimates  $P(S_{k,i})$ in (\ref{conditionalP}) instead of sampling proportional to these probabilities in the straightforward sampling scheme gives more precise estimates of $P(s_{k,i})$.

There are various choices of the number of samples in a stratum. One way is allocating samples according to their probabilities. That is, the number of samples in each stratum is proportional to its probability $P(S_{k,i})$. 
If a list needs 3000 contingencies, then we would have $3000 \times P(S_{2,1}) = 1748$ \includegraphics[height=9pt]{s21.pdf}, $3000 \times P(S_{3,1}) = 420$ \includegraphics[height=9pt]{s31.pdf}, $3000 \times P(S_{3,4}) = 18$ \includegraphics[height=9pt]{s34.pdf}, and 814 $s_*$. 
As shown in Section \ref{sec:sampleA}-A, this list  accounts for 70\% of outages in the test data. 
For strata of the three motifs, we randomly sample contingencies uniformly according to the fourth step of the straightforward sampling; for the stratum $S_{*}$, we can use the full straightforward sampling but exclude the previous three motifs when evaluating conditional probabilities in (\ref{equ:ps1}). 

The flexibility of the stratified sampling allows us to give more consideration to other factors. We may sample more contingencies from strata that we are interested in, or sample more contingencies from strata that generally have a high impact. 
For example, we could sample more contingencies from N-3 because N-3 generally has a higher impact than N-2. Stratified sampling reduces the variance of the estimate  when there are more samples in the stratum with high probability than with low probability when the stratum variance is the same \cite[Ch. 5.5]{cochran77}. 

Another example of the flexibility of stratified sampling is that it could be used in future work to better estimate risk.
%Risk evaluation is out of the scope of this paper; however, it is a direct use case. 
To evaluate the risk, the expected impact of contingencies in each stratum can be estimated by the average impact of contingency samples in that stratum, and the system risk is the weighted sum of the strata impacts with stratum probabilities as weights. 
Stratified sampling can reduce the variance of the risk estimates because contingencies are more homogeneous in each stratum than between strata, fewer samples are needed to obtain a precise estimate for each stratum, and combining these estimates for the whole population can be less variable.

\subsection{Deterministic contingency list}
A fixed or deterministic contingency list that involves no sampling could also be used.
As shown in Table~\ref{tbl:nkpatterns}, there are 2116 $s_{2,1}$ \includegraphics[height=9pt]{s21.pdf}, 4653 $s_{3,1}$ \includegraphics[height=9pt]{s31.pdf}, and 62 $s_{3,4}$ \includegraphics[height=9pt]{s34.pdf}. The total number of these three motifs is 6831, which can be simulated in an acceptable time. %As their probability is high, the stratified sampling prefers to sample all in each stratum to gain a better estimate with low variance.
Therefore we can make a contingency list that samples all the contingencies in the three motifs.
%If we only use the three contingency motifs as strata in the stratified sampling, the contingency sample is fixed, including all contingencies in the three motifs. 
This neglects the $S_*$ contingencies, but gives a deterministic contingency list that accounts for 78\% of multiple outages in the test data. 

\section{Test results on a second system}
This section applies the analysis of the previous sections to a second transmission system with outages recorded by the New York Independent System Operator (NYISO) and summarizes the results, which turn out to be similar. 

The NYISO transmission system outage records cover New York State and parts of neighboring US states and Canadian provinces, with more network detail in New York State.
The NYISO outage data is publicly accessed from its website \cite{nyisodata} and processed according to Carrington's method in \cite{nichelle21NAPS}. 
Twelve years of data are used here, spanning from 2008 to 2020. 
The NYISO network formed from the outage data using the method of \cite{DobsonPS16} has 1695 lines. The outage data do not have enough samples of N-4. Therefore, only N-2 and N-3 are considered. 
The first 10 years of data are used for training and the last 2 years of data are used for testing.

\looseness=-1
The contingency motifs identified in  the NYISO data are $S_{2,1}$ \includegraphics[height=9pt]{s21.pdf}, $S_{3,1}$ \includegraphics[height=9pt]{s31.pdf}, $S_{3,3}$ \includegraphics[height=9pt]{s33.pdf} and $S_{3,4}$ \includegraphics[height=9pt]{s34.pdf}, which are the same 2-edge and 3-edge motifs as in the BPA data. 
Then, we form the probabilistic model and sample contingencies according to the straightforward sampling scheme. 10\,000 contingencies turn out to cover 74\% of outages in the test data (0.9\% standard deviation). This shows that these contingency motifs also capture the spatial characteristics of the NYISO multiple initial outages.  
 
A contingency list is formed by the stratified sampling scheme. Motifs \includegraphics[height=9pt]{s21.pdf},  \includegraphics[height=9pt]{s31.pdf}, \includegraphics[height=9pt]{s34.pdf}, and all remaining patterns $S_*$ compose four strata.  
The consideration is the same as the BPA case: any individual subgraph in the three motifs has a higher probability than others; these three motifs already account for the most probability (75\%) of multiple outages; and the remaining motif \includegraphics[height=9pt]{s33.pdf} have a large number (19\,911) of distinct subgraphs (moreover, there are 6305 $s_{2,1}$, 14138 $s_{3,1}$, and 247 $s_{3,4}$).   
Since the NYISO network has more transmission lines than the BPA network, we form a contingency list with 10\,000 contingencies for demonstration. Allocating samples proportional to probabilities, this list contains 6305 $s_{2,1}$, 809 $s_{3,1}$, 247 $s_{3,4}$ and 2639 $s_{*}$, which accounts for 72\% of multiple initial outages in the test data.
%a list of 4000 contingencies has 2824 $s_{2,1}$, 324 $s_{3,1}$, 24 $s_{3,4}$, and 828 $s_{*}$. This list accounts for 51\% of outages in the test data.  

%\section{Conclusion and discussion}
\section{Discussion}
The contingency motifs could confirm or inform industry contingency selection by
indicating multiple contingencies with high probabilities. Given motifs in a power network, engineers with field knowledge can better identify vulnerable locations in the network for further analysis. In practical contingency analysis, contingencies selected from the contingency motifs could be refined by incorporating engineering considerations such as substation bus configurations. 

Because transmission line automatic outage data is sparse, it is routine to group together lines by characteristics such as voltage rating and component type to determine outage probabilities of the lines in the group.
The grouping of multiple initial line outages into contingency motifs is similar, except that the groups are formed by a  spatial pattern the multiple outages share, and it is a similarly pragmatic way to mitigate the sparsity of multiple line outage data when estimating the probabilities of multiple initial outages.

%This indicates a new way of grouping multiple contingencies to mitigate the limited outage data problem. Compared to always used grouping methods, such as voltage ratings, lengths, and component types, it uses more information from the outage spatial dependency. 

\looseness=-1
The method of this paper relies on a single database of detailed historical outage data that is routinely collected by transmission utilities in North America and also by many utilities worldwide.
The initial multiple outages can be readily extracted from the data and located on the network to find the contingency motifs and estimate their probabilities, so that the most probable contingencies can be sampled first. The network can be obtained from an inventory associated with outage data or, as we do in this paper, from the outage data itself.
Since their own data is available to each utility and the computations are not difficult, we would first recommend that contingency motifs be found and applied with specific utility data to get improved contingency lists to estimate cascading risk from simulations.
However, specific utility historical outage data may not be available, since the simulated system may not have associated historical  data, or the study is done outside a utility.  Then the risk assessment could still realize some similar benefits by relying on the overall similarities in power transmission system physics and engineering to use the contingency motifs 
\includegraphics[height=9pt]{s21.pdf},  \includegraphics[height=9pt]{s31.pdf}, and \includegraphics[height=9pt]{s34.pdf} 
that we observed in two different transmission systems.

This paper proposes an improved method of sampling multiple initial line outages, which is only one part of assessing cascading risk. 
We briefly comment on how the new method fits into assessing cascading risk. 
The new method can simply be substituted for the various sampling schemes with explicit or implicit uniform probability assumptions, and then combined with the sampling of the grid conditions and stresses, and the sampling of any cascading outages and interactions that may occur after the initial line outages to evaluate a cascading outcome. 
We note that for risk analysis it is necessary to sample likely as well as less likely possibilities across as much of the sample space as possible. 
For comprehensive reviews of cascading risk assessment we refer the reader to \cite{TFPESGM08,camsPS12}. Note that \cite{bialekPS16} explains the close relation between deterministic and probabilistic framings of cascading.

One feature of the new method that arises from it being directly driven by observed line outage data is that the initial line outage underlying causes need not be modeled and analyzed. 
The pattern of initial line outages in a given motif can arise from a variety of causes and mechanisms, but for the purpose of sampling initial line outages better, we only need to know the observed outcome of all these causes and mechanisms as it is expressed in the frequency of the motif. 
This point is specific to sampling initial outages for risk analysis, and we are not suggesting in any general way that the underlying causes are unimportant; indeed the underlying causes are vital to good engineering to mitigate the risk of specific situations.

There are some specific threats to grid security that can have atypical patterns of initial outages related to the specific threat, such as terrorism, war, and solar geomagnetic disturbances. 
In a grid in which one of these threats is rare, these atypical patterns will be rare in the historical record, and the motifs extracted from the historical record will tend not to include these atypical patterns. 
The extracted motifs correctly summarize the historical probability structure of the multiple outages in that grid and are appropriate for use in risk analysis on that basis.
On the other hand, one might choose to defend the grid against one of these threats that has rarely occurred in the past,
and in that case we would suggest that each of these threats requires separate analysis with their own initial contingency lists.
However, our improved contingency lists for multiple initial outages based on historical data can be expected to have some amount of 
overlap with these special contingency lists for two reasons.  First, in general, better screening of multiple initial outages is broadly helpful because all threats share the same 
grid topology and protection systems.  
Second, in particular, many motifs correspond to outages related to a single substation, and 
efficiently augmenting the contingency analysis 
with these motifs tends to cover some of the starting impact of these threats
if that threat starts with damage to a single substation.
Earthquakes are another threat that may require separate analysis because they tend to have an 
unusually large number of initial outages occurring in the same minute in a specific area of the grid, but can be rare in any 
given specific area, giving only a few cases in recorded data.

%Contingency motifs can substantially improve the contingency lists and the risk estimates obtained when assessing cascading risk with respect to N-k contingencies with simulations.

%Outage data are needed for a different power system to detect contingency motifs. On the other hand, since the basic structures of power systems, such as bus configurations, transmission lines on a right-of-way, or protection control groups, are common across different systems, we suggest starting with contingencies of \includegraphics[height=9pt]{s21.pdf} ,  \includegraphics[height=9pt]{s31.pdf} , and \includegraphics[height=9pt]{s34.pdf} when outage data are not available.

\looseness-1
In implementation, we expect the new sampling methods to be used offline to generate a fixed contingency list. Then it takes the same time online or offline as any other contingency list, which is a small part of the time required for cascading simulation.

\section{Conclusion}
In going beyond  N-1 security,
contingency lists of  multiple initial line outages are foundational for assessing cascading risk and the security of power transmission systems.
We analyze the spatial patterns of multiple automatic line outages that occurred in the same minute at the start of a cascade from historical outage data. 
Some patterns occur significantly more frequently in outage data than in random subgraphs of the particular power network under consideration. 
We call these patterns contingency motifs. 
The existence of contingency motifs is the result of complex physical and engineering dependencies in power systems. 

Three contingency motifs (\includegraphics[height=9pt]{s21.pdf}, \includegraphics[height=9pt]{s31.pdf}, and \includegraphics[height=9pt]{s34.pdf}) account for most of the probability of multiple line outages in both BPA and NYISO historical data. 
A contingency list formed from these contingency motifs is much more efficient than random selection or exhaustive listing. 
This improved contingency list  can easily improve simulations that evaluate cascading risk. 
Specifically, to assess the cascading risk, we can use all contingencies of the contingency motifs or sample a desired number of multiple initial outages according to a sampling scheme. 
A stratified sampling scheme is flexible and  effective.

We show that sampling based on motifs works on the outage datasets of two transmission systems.
We expect that it will also be applicable to other transmission systems, and we hope that others will be able to confirm this with the outage data available to them.

\looseness-1
Contingency motifs can substantially improve the contingency lists and the risk estimates obtained when assessing cascading risk with respect to N-k contingencies with simulations.

\section{Acknowledgments}
The authors gratefully thank Bonneville Power Administration and New York Independent System Operator for making publicly available the data on which this paper is based.
 The analysis and any conclusions are strictly those of the authors and not of BPA or NYISO.
Support from NSF grant 2153163 is gratefully acknowledged.

\appendix[Detecting contingency motifs]
\label{app1}
\subsection{Statistical model of multiple contingencies}
Let $X$ be the number of $s_{k,i}$ in $n_k$ N-ks. For simplicity, we write $P(S_{k,i}|k)$ as $p_{ki}$ and $P^{\rm uni}(S_{k,i}|k)$ as $p^{\rm uni}_{ki}$.
$X$ follows a binomial distribution:
\begin{align}
    P_{k}(X = x) = \binom{n_k}{x} p_{{ki}}^{x} \left( 1-p_{{ki}} \right)^{n_k - x }
\end{align}
\subsection{Frequentist hypothesis test}
Under $H_0$, the likelihood of obtaining $n_{k,i}$ or more $s_{k,i}$ is 
\begin{align}
    L \left(p_{{ki} }|n_{k,i} \right) = \sum_{j=n_{k,i}}^{n_{k}} \binom{n_k}{j} (10 p_{{ki}}^{\rm uni})^j (1-10 p_{{ki}}^{\rm uni})^{n_{k} - j }
\end{align}
When the likelihood is less than significance level 0.01, we reject $H_0$, which means that the probability that $H_0$ is true but we reject it is less than 0.01.

\subsection{Bayesian hypothesis test}
We compare the posterior probability $P(H_0 | n_{k,i})$ with $P(H_1 | n_{k,i})$. If $P(H_0 | n_{k,i}) \geq P(H_1 | n_{k,i})$, we accept $H_0$; otherwise, we reject $H_0$ and accept $H_1$.
\begin{align}
    P(H_0 | n_{k,i}) & = \frac{P(n_{k,i} | H_0) P(H_0)}{P(n_{k,i})} \notag\\
    & = \frac{P(n_{k,i} | H_0) P(H_0)}{P(n_{k,i} | H_0) P(H_0) + P(n_{k,i} | H_1) P(H_1)} \notag\\
%    & = \frac{1}{1 + \frac{P(n_{k,i} | H_1) P(H_1)}{P(n_{k,i} | H_0) P(H_0)}}
& =1\Big/\Big(1 + \frac{\scriptstyle P(n_{k,i} | H_1) P(H_1)}{\scriptstyle P(n_{k,i} | H_0) P(H_0)}\Big)
\end{align}
where $P(H_0)$ is the prior probability and 
\begin{align}
    P(n_{k,i} | H_0) = \int_{0}^{10 p_{{ki}}^{\rm uni}} P(n_{k,i} | p_{{ki}}) f(p_{{ki}} | H_0) dp_{{ki}}
\end{align}
is the marginal likelihood under $H_0$. $f(p_{ki} | H_0)$ is the prior for parameter $p_{ki}$ when $H_0$ is true, which is assumed to be a uniform distribution. 

Assume $P(H_0) = P(H_1) =0.5$, and $p_{ki} | H_0$ and $p_{ki} | H_1$ both follow uniform distributions. Then the posterior probability of $H_0$ is
\vspace{-4mm}
\begin{align}
    P(H_0 | n_{k,i}) = \frac{1}{1+BF}
\end{align}
where BF is the Bayes Factor for $H_1$ relative to $H_0$, which can be evaluated as follows:
%
%
%(\ref{equ:bf}) shows the derivation of the Bayesian Factor. 
%
\begingroup
\allowdisplaybreaks
\begin{align}
    BF&  = \frac{P(n_{k,i} | H_1)}{P(n_{k,i} | H_0)} \notag\\
    =&  \frac{\int_{10 p_{ki}^{\rm uni}}^{1} P(n_{k,i} | p_{ki}) f(p_{ki} | H_1) dp_{ki}}{\int_{0}^{10 p_{ki}^{\rm uni}} P(n_{k,i} | p_{ki}) f(p_{ki} | H_0) dp_{ki}} \notag \\
    =&   \frac{ \int_{10 p_{ki}^{\rm uni}}^{1}  p_{ki}^{n_{k,i}} \left( 1-p_{ki} \right)^{ n_{k} - n_{k,i} } dp_{ki} }{\int_{0}^{10 p_{ki}^{\rm uni}} p_{ki}^{n_{k,i}} \left( 1-p_{ki} \right)^{n_{k} - n_{k,i} } dp_{ki}}
    = \frac{1-F(10 p_{ki}^{\rm uni})}{F(10 p_{ki}^{\rm uni})} \notag
\end{align}
\endgroup
where $F(x)$ is the cumulative density function of a beta distribution with parameters $n_{k,i}+1$ and $n_{k} - n_{k,i} + 1$. 

\bibliographystyle{IEEEtran}
\bibliography{IEEEabrv,references}

\begin{IEEEbiographynophoto}{Kai Zhou}
(Member, IEEE) received the B.S degree in electrical engineering from China Agriculture University in 2014, the M.S. degree in electrical engineering from Tianjin University in 2017, and Ph.D. degree in electrical engineering from Iowa State University in 2022. He is currently an assistant professor at Soochow University, Suzhou, China. His research interests include cascading failures, resilience, and risk analysis.
\end{IEEEbiographynophoto}

\begin{IEEEbiographynophoto}{Ian Dobson}
(Fellow, IEEE) received the B.A. degree in mathematics from Cambridge University, UK, and the Ph.D. degree in electrical engineering from Cornell University, USA.
He is currently Sandbulte Professor of Engineering at Iowa State University, Ames, IA, USA.
\end{IEEEbiographynophoto}

\begin{IEEEbiographynophoto}{Zhaoyu Wang}
(Senior Member, IEEE) received the B.S. and M.S. degrees in electrical engineering from Shanghai Jiaotong University, and the M.S. and Ph.D. degrees in electrical and computer engineering from Georgia Institute of Technology. He is the Northrop Grumman Endowed Associate Professor with Iowa State University. His research interests include optimization and data analytics in power distribution systems and microgrids. He was the recipient of the National Science Foundation CAREER Award, the Society-Level Outstanding Young Engineer Award from IEEE Power and Energy Society (PES), the Northrop Grumman Endowment, College of Engineering’s Early Achievement in Research Award, and the Harpole-Pentair Young Faculty Award Endowment. He is the Principal Investigator for a multitude of projects funded by the National Science Foundation, the Department of Energy, National Laboratories, PSERC, and Iowa Economic Development Authority. He is the Co-TCPC of IEEE PES PSOPE, the Chair of IEEE PES PSOPE Award Subcommittee, the Vice Chair of PES Distribution System Operation and Planning Subcommittee, and the Vice Chair of PES Task Force on Advances in Natural Disaster Mitigation Methods. He is an Associate Editor of IEEE TRANSACTIONS ON SUSTAINABLE ENERGY, IEEE OPEN ACCESS JOURNAL OF POWER AND ENERGY, IEEE POWER ENGINEERING LETTERS, and IET Smart Grid. He was an Associate Editor for IEEE TRANSACTIONS ON POWER SYSTEMS and IEEE TRANSACTIONS ON SMART GRID.
\end{IEEEbiographynophoto}
\vfill

\end{document}